%% file: main_second_submission.tex
\documentclass[
twocolumn,
prl,
superscriptaddress,
nofootinbib,
showpacs
]{revtex4-1} 

\usepackage{graphicx}
\usepackage{dcolumn}
\usepackage{amsmath}
\usepackage{amssymb}
\usepackage{epstopdf}

\makeatletter
\def\btt#1{\texttt{\@backslashchar#1}}%
\DeclareRobustCommand\bblash{\btt{\@backslashchar}}%
\makeatother


\begin{document}
\title{First Indication of Terrestrial Matter Effects on
Solar Neutrino Oscillation}
\input{authors}
\date{\today}

\begin{abstract}
We report an indication that the elastic scattering rate of solar
$^8$B neutrinos with electrons in the Super-Kamiokande detector is larger
when the neutrinos pass through the Earth during nighttime. We determine
the day/night asymmetry, defined as the difference of the average day rate
and average night rate divided by the average of those two rates, to
be $(-3.2\pm1.1(\text{stat})\pm0.5(\text{syst}))\%$, which deviates from zero
by 2.7 $\sigma$. Since the elastic scattering process
is mostly sensitive to electron-flavored solar neutrinos, a non-zero
day/night asymmetry implies that the flavor oscillations of solar neutrinos
are affected by the presence of matter within the neutrinos' flight path.
Super-Kamiokande's day/night asymmetry is consistent with neutrino
oscillations for $4\times10^{-5}$eV$^2\leq\Delta m^2_{21}\leq7\times10^{-5}$eV$^2$ and large mixing values of $\theta_{12}$, at the $68\%$ C.L.
\end{abstract}


\maketitle

Neutrino flavor oscillations occur when the phase difference of a superposition
of massive neutrinos changes. Such phase changes occur while neutrinos
are propagating in vacuum (vacuum oscillations).
Wolfenstein~\cite{w} realized that the neutrino-electron elastic
forward-scattering amplitude introduces additional
phase shifts. As a consequence, neutrinos propagating in matter will
oscillate differently than neutrinos propagating through vacuum. These
matter effects are a fundamental prediction of the present theory of
neutrino oscillations. In this letter, we report
an indication of the existence of such matter effects.

Vacuum oscillations cannot easily explain a solar neutrino electron-flavor
survival probability $P_{ee}$ which is measured to be significantly below one
half~\cite{davis,kamiokande,sk1,sk2,sk3,firstsno,sno}, in the energy region of
$\sim$8 to 18 MeV. Mikheyev and Smirnov~\cite{ms} explained this
experimental fact as the adiabatic transformation of the neutrinos through the
varying solar density causing a resonant conversion to the second mass
eigenstate inside the Sun (MSW resonance). Lower energy ($<$2 MeV) solar
neutrino data~\cite{othersolar} are still described  well as averaged vacuum
oscillations. However, searches for the transition of $P_{ee}$ from
the MSW resonance to vacuum oscillations (near 3 MeV) were so far
unsuccessful~\cite{sno,sk4}.  Moreover, these previous observations
imply matter effects only indirectly, since there is no ``control beam'' of 
solar neutrinos that only propagates in vacuum.
Atmospheric neutrino experiments can probe the
existence of matter effects within the Earth in a similar fashion, and while
there is currently no significant departure of present atmospheric
data~\cite{skatm1,skatm} from the vacuum oscillation predictions, these
effects will serve to determine the mass hierarchy and CP
phase in future atmospheric and long baseline experiments.

The cleanest and most direct test of matter effects on neutrino oscillations
is the comparison of the daytime and the nighttime solar neutrino interaction
rates (solar day/night effect). In this comparison, the solar zenith angle
controls the size and length of the terrestrial matter density through
which the neutrinos pass, and thereby the oscillation probability and the
observed interaction rate. An increase in the nighttime interaction
rates implies a regeneration of electron-flavor neutrinos.
Other solar neutrino
measurements~\cite{kamiokande,sno,borexinodn} have found
no significant day/night differences.
Here, we report a 2.7 $\sigma$ indication of a non-zero
solar neutrino day/night effect.\footnote{This is consistent with the strong
limits of~\cite{borexinodn} using
861 keV mono-energetic neutrinos, an energy where presently-preferred
neutrino oscillation parameters predict no day/night effect.}

Super-Kamiokande (SK) is a 50,000 metric ton cylindrical water Cherenkov
detector. The optically separated 32,000 ton inner detector (ID),
viewed by $\sim$11,100 50 cm diameter photomultiplier tubes (PMTs), is
surrounded by an 18,000 ton active veto shield, viewed by $\sim$1,900 20 cm
diameter PMTs. The detector is described in detail in~\cite{sk_nim}. SK
detects recoil electrons coming from the
elastic scattering of solar neutrinos with electrons. Only $^8$B and $hep$
solar neutrinos produce recoil electrons of sufficiently high energy
to be detected in SK. Neutrino-electron elastic scattering is mostly sensitive
to electron-flavored neutrinos, because the cross section for
$\nu_{\mu,\tau}$ scattering is six times smaller, since only the electroweak
neutral-current interaction channel contributes. The scattering vertex is 
reconstructed using the timing of the Cherenkov light, while the direction and 
energy of
the recoil electrons are determined from the light pattern and intensity.
For 10 MeV electrons in SK-IV, the vertex resolution is 52 cm, the directional
resolution is 25$^\circ$ (limited by multiple Coulomb scattering), and the
energy resolution is 14.0$\%$ (dominated by Poisson fluctuations of the number
of photons detected with $\sim$6 photo-electrons per MeV). More details
are given in~\cite{sk1,sk2,sk3,sk4}.

There are four distinct phases of SK.
Initially, SK-I had 11,146 ID PMTs and used 1,496 live days between 1996 and
2001 for low-energy analysis \cite{sk1}. In 2001, an accident
during maintenance destroyed $\sim$7,000 ID PMTs.  5,182 surviving and spare ID
PMTs were then deployed for SK-II \cite{sk2}, running between 2002-2005, with a
total low-energy analysis livetime of 791 days.  After full reconstruction of
the detector, SK-III \cite{sk3} took data between 2005 and 2008, acquiring
548 live days for the low-energy analysis, with 11,129 ID PMTs.
Since 2008, SK-IV has been running with upgraded electronics and
data acquisition system (DAQ).  In this paper, data from SK-I, SK-II, SK-III
and 1,306 live days of SK-IV are used \cite{sk4}. The day/night analysis uses
recoil electrons above 4.5 MeV in SK-I, III, and IV and 7 MeV in SK-II.
The energy range of the SK-I-III day/night analysis is identical as in
~\cite{sk1,sk2,sk3}; however, here it is quoted as kinetic energy
(we simply subtracted 0.5 MeV) while \cite{sk1,sk2,sk3} use total energy.
In ~\cite{sk4}, the observed solar neutrino signal at lower recoil electron
energies is used for the flux and spectrum analysis.

At the time of each event, the solar zenith angle $\theta_z$ is determined.
This is the angle between the vector from the solar position to the event
vertex and the vertical detector ($z$) axis. The precision of the cosine of
this angle is much better than $10^{-3}$, the bin width used in the following
analysis.
The accuracy of this angle is limited only by SK's absolute time precision
(a few 100 ns) and basic astronomy. The SK elastic scattering rate as a
function of the solar zenith angle $r(\cos\theta_z)$ is used to search for
a day/night difference in the interaction rate. The expected change in the
interaction rate due to the varying Sun-Earth distance (induced by the
eccentricity of Earth's orbit) is taken into account throughout this paper.
The most straight-forward method to look for a day/night effect is to
define separate day ($\cos\theta_z\leq0$) and night ($\cos\theta_z>0$)
samples. Based on $r_D$ ($r_N$), the average scattering rate of the day
(night) sample, we define the SK day/night
asymmetry as $A_{\text{DN}}=(r_D-r_N)/\frac{1}{2}(r_D+r_N$).  Therefore,
$A_{\text{DN}}=0$ implies no terrestrial matter effect on solar neutrino
oscillations.

To increase sensitivity, \cite{dn} first introduced an unbinned maximum
likelihood fit of the
solar zenith angle distribution of the rate $r(\cos\theta_z)$ to the
day/night variation amplitude $\alpha$. This was done using ``shapes'' of such
variations expected from neutrino oscillation calculations.
By construction, $\alpha$ scales the calculated day/night
asymmetry $A_{\text{DN}}^{\text{calc}}$ while leaving the average rate
unchanged, giving the measured day/night asymmetry
$A_{\text{DN}}^{\text{fit}}=A_{\text{DN}}^{\text{calc}}\times\alpha$.
This more sophisticated method, referred to as the ``amplitude fit'', was also
used in~\cite{sk1}.  The calculated oscillation shapes
ignore the SK daytime overburden, which can be up to a few kilometers
depending on the solar zenith angle.

We refer to the angle between the solar and reconstructed recoil electron
candidate directions as $\theta_{\text{sun}}$. The solar neutrino interaction
rate is extracted by an extended maximum likelihood fit~\cite{sk1} to the
$\cos\theta_{\text{sun}}$ distribution.  \cite{dn} expands the signal
likelihood to allow
for a time-dependent solar neutrino-electron elastic scattering rate,
parameterized by the amplitude scaling variable $\alpha$. The best-fit
$\alpha$, multiplied with $A_{\text{DN}}^{\text{calc}}$, defines
a best-fit $A_{\text{DN}}^{\text{fit}}$. In this manner
the day/night asymmetry is measured more precisely statistically. It is also
less vulnerable to some key systematic effects, such as directional variation
of the energy scale (the frequency of which is limited by SK's angular
resolution, $\sim25^\circ$).

Because the amplitude fit depends on the shape of the day/night
variation, given for each energy bin in~\cite{dn} (and also in \cite{sk1}), it
necessarily depends on the assumed oscillation parameters.
Vacuum oscillations depend on the neutrino energy, the length of the flight
path $L$ and the oscillation parameters; the difference of the squared masses
of the mass eigenstates $\Delta m^2_{ij}$ ($i=1,2,3,...$) and
the mixing of the mass eigenstates with the flavor eigenstates (mixing angles
$\theta_{ij}$).  If the neutrinos
propagate through matter, then the density of the matter will effectively
change the oscillation parameters.  It was shown analytically
by~\cite{smirnov,blennow,akhmedov} that
$r(\cos\theta_z)-r_D$ oscillates with the vacuum frequency
$\frac{\Delta m^2_{21}}{4E}L$, if $L\approx2R\cos\theta_z$ denotes now the
path length of the neutrino inside the Earth ($R$ being the Earth radius).
Although the dependence of matter effects on the mixing angles (in or near the
large mixing angle solutions and for $\theta_{13}$ values consistent with
reactor neutrino measurements~\cite{reactorexp}) is quite small,
the dependence on $\Delta m^2_{21}$ is more noticeable.
The fit is run for solar oscillation
parameter sets which predict various matter effects
($10^{-9}$ eV$^2\le\Delta{m_{21}^2}\le10^{-3}$ eV$^2$ and $10^{-4}\le\sin^2\theta_{12}\le1$), for values of $\sin^2\theta_{13}$ between 0.015 and 0.035.

\begin{figure}
   \includegraphics[width=8cm]{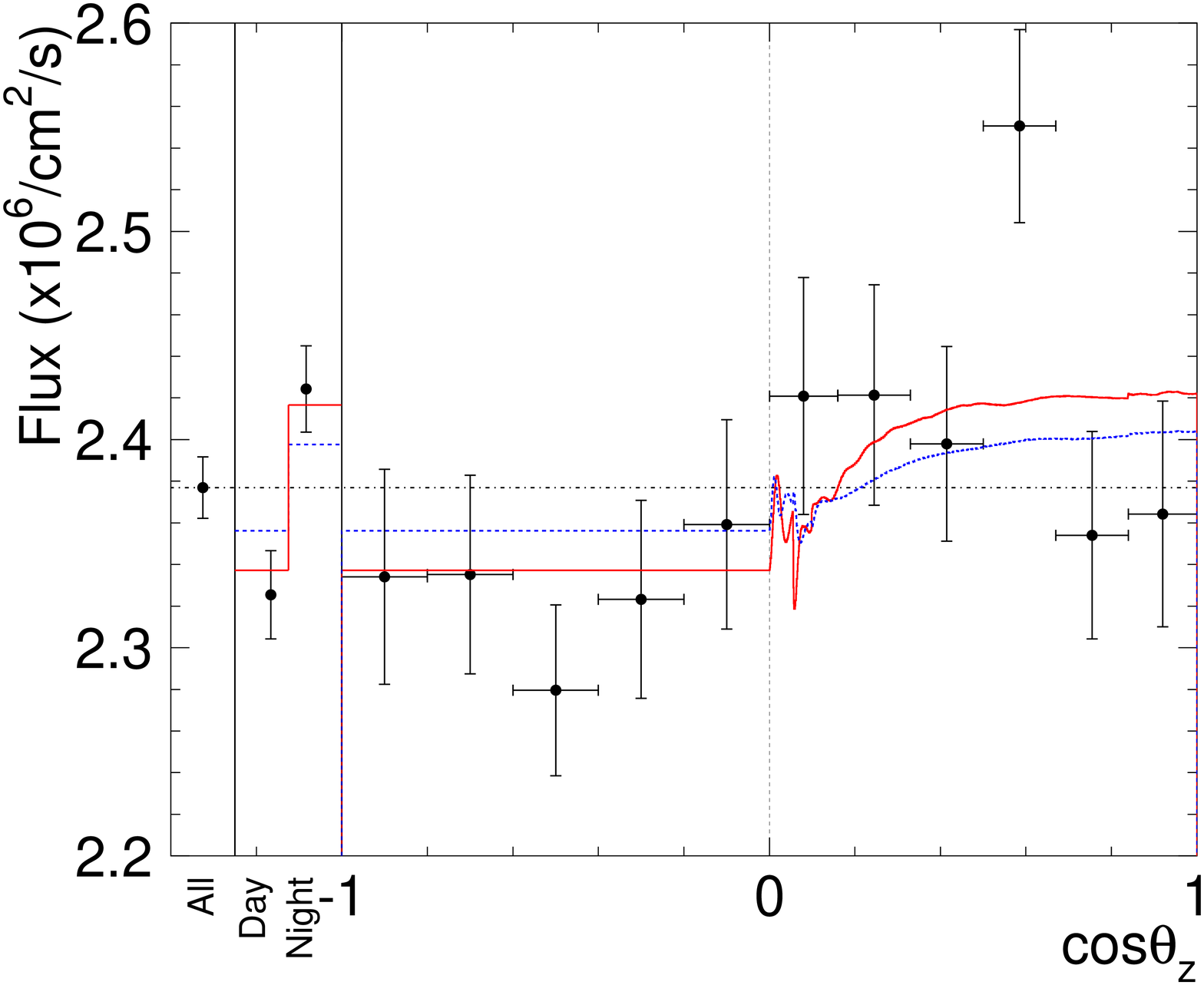}
    \caption{SK combined solar zenith angle dependence of the $^8$B
solar neutrino flux.  Solid red (dashed blue) gives the prediction based on
oscillation parameters from a fit to SK data while constraining the flux
(solar+KamLAND fit) and the dashed-dotted line gives the total average flux.
\label{fig:dnall}}
\end{figure}

Fig.~\ref{fig:dnall} combines the data from all four SK phases to show
the measured zenith angle distribution of the flux assuming no oscillations.
The expected zenith variation assuming
best-fit oscillation parameters~\cite{sk4} from a global fit based on
solar neutrino data~\cite{sk1,sk2,sk3,sk4,sno,davis,othersolar}  is overlaid
in solid red.  The dashed blue line also includes reactor anti-neutrino
data~\cite{kamland}.
The day and night flux values given in the left portion of
Fig.~\ref{fig:dnall}
imply a day/night asymmetry of $A_{\text{DN}}=(-4.2\pm1.2(\text{stat}))\%$.
The ninth data point in the right panel of Fig.~\ref{fig:dnall} shows an
$\sim2$ sigma deviation, enhancing the resulting $A_{DN}$.  We believe this is
a statistical flucuation, accounted for by the quoted statistical
uncertainties.  The result $A^{\text{fit}}_{DN}$ coming from the unbinned
maximum likelihood fit is less prone to these types of flucuations.
To calculate the total systematic uncertainty, the individual systematic
uncertainties of the four phases (for values see~\cite{sk1,sk2,sk3,sk4}) are
assumed
to be uncorrelated, since the dominant contributions come from the energy-scale uncertainty (tuned independently for each phase) and the background
directional distribution shape uncertainty (evaluated from detector zenith
angle data distributions and limited by statistical fluctuations).
The measured day/night asymmetry when using this simple method is shown in the
middle column of Table~\ref{tab:DN}, along with the statistical and systematic
uncertainties.  SK measures the day/night asymmetry in this simple way as
$A_{\text{DN}}=(-4.2\pm1.2(\text{stat})\pm0.8(\text{syst}))\%$, which deviates
from zero by 2.8 $\sigma$.

\begin{table}[t]
  \caption{Day/night asymmetry for each SK phase, coming from separate day
and night rate measurements (middle column) and the amplitude fit (right
column). The uncertainties shown are statistical and systematic.
The entire right column assumes the SK best-fit point of oscillation
parameters.}
  \begin{tabular}{l c c}
    \hline\hline
                   &  $A_{\text{DN}}\pm(\text{stat})\pm(\text{syst})$ & $A_{\text{DN}}^{\text{fit}}\pm(\text{stat})\pm(\text{syst})$
    \\ \hline
    SK-I           & $(-2.1\pm2.0\pm1.3)\%$ & $(-2.0\pm1.7\pm1.0)\%$ \\
    SK-II          & $(-5.5\pm4.2\pm3.7)\%$ & $(-4.3\pm3.8\pm1.0)\%$ \\
    SK-III         & $(-5.9\pm3.2\pm1.3)\%$ & $(-4.3\pm2.7\pm0.7)\%$ \\
    SK-IV          & $(-5.3\pm2.0\pm1.4)\%$ & $(-3.4\pm1.8\pm0.6)\%$ \\ \hline
    Combined       & $(-4.2\pm1.2\pm0.8)\%$ & $(-3.2\pm1.1\pm0.5)\%$ \\
    \hline\hline
  \end{tabular}
  \label{tab:DN}
\end{table}

\begin{figure}[t]
   \includegraphics[width=8cm]{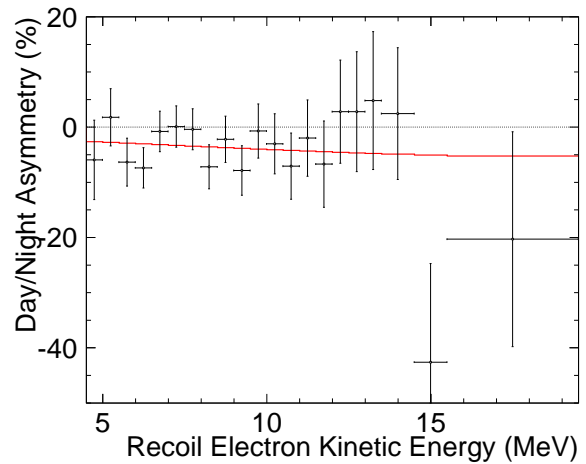}
    \caption{SK day/night amplitude fit as a function of recoil electron
kinetic energy (unlike~\cite{dn} which uses total energy),
shown as the measured amplitude times the expected day/night asymmtery,
for oscillation parameters chosen by the SK best-fit.  The error bars
shown are statistical uncertainties only and the expected dependence is shown
in red.
\label{fig:dnspectrum}}
\end{figure}

Fig.~\ref{fig:dnspectrum} shows the combined
SK-I/II/III/IV day/night amplitude fit as a function of recoil electron
energy.  In each recoil electron energy bin $e$, the day/night variation
is fit to an amplitude $\alpha_e$. The displayed day/night asymmetry
values are the product of the fit amplitude $\alpha_e$ with the expected
day/night asymmetry $A_{\text{DN},\text{calc}}^e$ (red), when using the SK
best-fit point of oscillation parameters
($\Delta{m_{21}^2}=4.8^{+1.8}_{-0.9}\times10^{-5}$ eV$^2$,
$\sin^2\theta_{12}=0.342^{+0.029}_{-0.025}$ \cite{sk4} and
$\sin^2\theta_{13}=0.025\pm0.003$ \cite{reactorexp}).
These parameters are
chosen when using SK's spectral and time variation data along with constraints
on the $^8$B solar neutrino flux and $\theta_{13}$.
When all energy bins are fit together and the same oscillation
parameters assumed, the resulting SK-measured day/night asymmetry
coming from the amplitude fit is
$A_{\text{DN}}^{\text{fit}}=(-3.2\pm1.1(\text{stat}))\%$, with an asymmetry of
$-3.3\%$ expected by numerical calculations (see \cite{dn} for details).

\begin{table}[b]
  \caption{Day/night amplitude fit systematic uncertainties by SK phase.  The
total is found by adding the contributions for each phase in quadrature.}
  \begin{tabular}{l c c c c}
  \hline\hline
                    & SK-I    & SK-II   & SK-III  & SK-IV    \\ \hline
  Energy Scale      & 0.8$\%$ & 0.8$\%$ & 0.2$\%$ & 0.05$\%$ \\
  Energy Resolution & 0.05$\%$ & 0.05$\%$ & 0.05$\%$ & 0.05$\%$  \\
  Background Shape  & 0.6$\%$ & 0.6$\%$ & 0.6$\%$ & 0.6$\%$  \\
  Event Selection   & ---     & ---     & 0.2$\%$ & 0.1$\%$  \\
  Earth Model~\cite{prem} & 0.01$\%$ & 0.01$\%$ & 0.01$\%$ & 0.01$\%$  \\ \hline
  Total             & 1.0$\%$ & 1.0$\%$ & 0.7$\%$ & 0.6$\%$  \\
  \hline\hline
  \end{tabular}
  \label{tab:DNsys}
\end{table}
Originally the systematic uncertainties on the SK-I and II day/night amplitude
measurements (see~\cite{dn}) were conservatively assigned
to be the same as that of the simple day/night asymmetry measurement
(see~\cite{sk1,sk2}).  Because \cite{sk1,sk2} only give total systematic
uncertainties and not those for each of the components, we have now
re-estimated the
systematic uncetainties of the day/night amplitude fit of the first two SK
phases, using similar methods as for SK-III and IV.
The methods for estimating the systematic uncertainties
of the amplitude fit in SK-III and IV are detailed in \cite{sk4} (see Section
9.3).  A summary of the various components of the systematic uncertainty on
the day/night amplitude fit, as well as the total, is given in
Table~\ref{tab:DNsys} for each SK phase.

During the SK-I and II phases, the largest contribution to
the systematic uncertainty came from the directional dependence of
the energy scale. From the beginning of the SK-III phase, a depth-dependent
water transparency parameter was introduced into the MC simulation program.
This corrects for the depth-dependence of the water absorption coefficient and
greatly reduces the directional dependence of the energy scale.  The further
reduction seen from SK-III to SK-IV comes from an improvement in the
comparison between data and MC timing, the result of the electronics upgrade
prior to SK-IV.  The largest
contribution to the systematic uncertainty now comes from the expected
background shapes, which are derived from fits to the detector's zenith and
azimuthal angle distributions after statistical subtraction of the solar
neutrinos.  The accuracy of these shapes are limited by statistics.

The additional contribution to the systematic uncertainty during SK-III and
IV, coming from the event selection, is the result of the combination of the
external event and tight fiducial volume cuts (see \cite{sk3,sk4}).  The tight
fiducial volume cut introduced at the start of SK-III is asymmetric in the $z$
direction, causing the external event cut to have different selection
efficiencies during the day and night times.  As for the case of the simple
day/night asymmetry
measurement, the total systematic uncertainty of each SK phase is assumed to
be uncorrelated, and is added in quadrature to the statistical uncertainty
of the corresponding phase before combining the results of each phase together.

The right column of Table~\ref{tab:DN} lists the measured day/night asymmetry
coming from the amplitude fit to each phase, as well as the combined fit,
for oscillations parameters at the SK best-fit
point. The combined fit takes into account energy threshold and resolution.
The equivalent SK day/night asymmetry coming from the amplitude fit is
\begin{equation*}
A_{\text{DN}}^{\text{fit}}=(-3.2\pm1.1\text{(stat)}\pm0.5\text{(syst)})\%,
\end{equation*}
\noindent which differs from zero by 2.7 $\sigma$.  The measured value of the
day/night asymmetry agrees with $-3.2\%$, within $\pm0.2\%$, for all sets of
oscillation parameters contained in the large mixing angle region.
The SNO experiment also performed a search for the day/night variation of
the charged-current interaction rate on deuterium \cite{sno}. From
their result, we predict the SK day/night asymmetry to be
$A_{\text{DN}}^{\text{pred}}=(-2.0\pm1.8)\%$\footnote{SNO actually models
the night/day asymmetry of the survival probability as
$a_0+a_1(E_{\nu}-10$ MeV$)$ and fits the coefficients $a_i$ \cite{sno}. We
scale
the expected coefficients (based on $\Delta m^2_{21}=4.84\times10^{-5}$eV$^2$)
by an amplitude $\alpha_{\text{SNO}}$ and minimize the SNO $\chi^2$ with
respect to it. SNO data then implies $a_0=(3.4\pm2.9)\%$, while SK and SNO
combined is equivalent to $a_0=(4.8\pm1.6)\%$}. Combining SK and SNO data
yields $A_{\text{DN}}^{\text{fit}}=(-2.9\pm1.0\mbox{(stat+syst)})\%$,
which differs from zero by 2.9 $\sigma$. The expected SK day/night asymmetry
for these oscillation parameters is $-3.3\%$.  Changing $\Delta m_{21}^2$
to $7.41\times10^{-5}$ eV$^2$ and $\sin^2\theta_{12}$ to 0.31 (motivated by
KamLAND data \cite{kamland}) changes the SK-measured day/night asymmetry to
$(-3.0\pm1.0(\text{stat})\pm0.5(\text{syst}))\%$, slightly reducing
the significance for a non-zero day/night asymmetry from 2.7 to 2.6 $\sigma$.

\begin{figure}[t]
  \begin{center}
    \includegraphics[trim=0cm 0cm 0cm 0cm,width=8cm,clip]{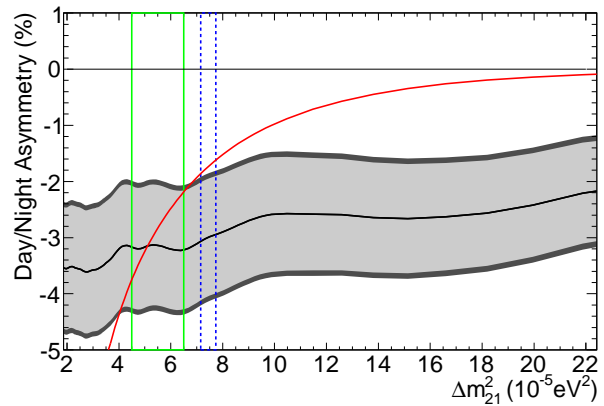}
    \caption{Dependence of the measured day/night asymmetry (fitted day/night
amplitude times the expected day/night asymmetry (red)) on $\Delta{m_{21}^2}$,
for $\sin^2\theta_{12}=0.314$ and $\sin^2\theta_{13}=0.025$. The 1 $\sigma$
statistical uncertainties are given by the light gray band.  The additional
dark gray
width to the band shows the inclusion of the systematic uncertainties.
Overlaid are the 1 $\sigma$ allowed ranges from the solar global fit
(solid green) and the KamLAND experiment (dashed blue). \label{fig:dndm2}}
  \end{center}
\end{figure}

Fig.~\ref{fig:dndm2} shows the $\Delta{m_{21}^2}$ dependence of the equivalent
day/night asymmetry of the SK combined amplitude fit for
$\sin^2\theta_{12}=0.314$ and $\sin^2\theta_{13}=0.025$. The expected day/night
asymmetry is indicated by the red curve. The point where the best-fit crosses
the expected curve represents the value of $\Delta{m_{21}^2}$ where the
measured day/night amplitude is $\alpha=1$. Superimposed are the 1 $\sigma$ 
allowed
ranges in $\Delta{m_{21}^2}$ from the solar global fit~\cite{sk4} (green) and
from the KamLAND experiment~\cite{kamland} (blue). The amplitude fit has
negligible
dependence on the values of $\theta_{12}$ (within the large mixing angle region
of oscillation parameters) and $\theta_{13}$
($0.015\leq\sin^2\theta_{13}\leq0.035$), leading to a $68\%$ C.L. allowed
range of $4\times10^{-5}$ eV$^2\leq\Delta m^2_{21}\leq7\times10^{-5}$ eV$^2$
(as shown in Fig.~\ref{fig:dndm2}).
Aside from the amplitude of the day/night asymmetry, another handle to
$\Delta m^2_{21}$ is the day/night variation frequency.
Although the amplitude of the day/night variation is too small (compared to
present uncertainties) to measure the frequency, some frequencies are favored
by about 2 $\sigma$ over others.

\begin{figure}[t]
  \begin{center}
    \includegraphics[width=8.5cm,clip]{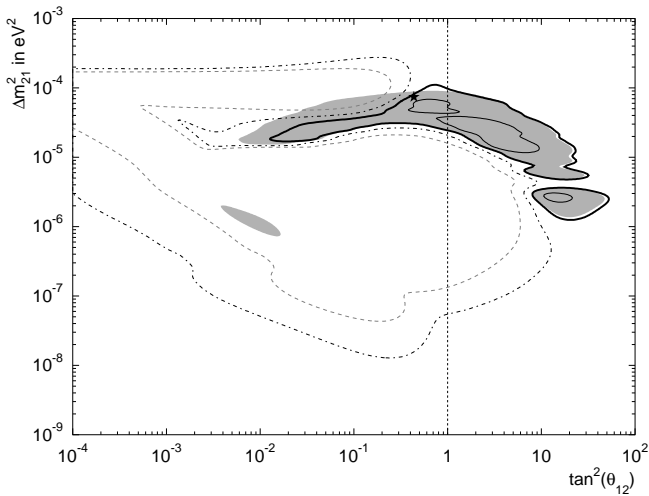}
    \caption{Contours at 68$\%$ (solid thin line), 95$\%$ (solid
thick line), three sigma (dashed-dotted line), and five sigma (dashed gray
line) of the flux-independent SK solar neutrino oscillation analysis.
The shaded gray area indicates regions preferred by the day/night variation
data for the 95$\%$ case.  The best-fit parameters resulting from a fit to all
solar neutrino \cite{sno,othersolar,sk4,homestake} and KamLAND \cite{kamland}
data is shown by the black asterisk.
 \label{fig:osccontour}}
  \end{center}
\end{figure}

Even so, the neutrino flux-independent solar neutrino oscillation analysis
of~\cite{dn} uses frequency and amplitude of the day/night variation as well
as spectral information. We calculate the log likelihood ratio
between $\alpha=1$ and $\alpha=0$, multiply by $-2$, and then add it to the
$\chi^2$ values of the fit to the recoil electron spectrum (see~\cite{sk4}).
Fig.~\ref{fig:osccontour} shows the flux-independent SK-I/II/III/IV
contours of 68$\%$ (solid thin line), 95$\%$ (solid
thick line), three sigma (dashed-dotted line), and five sigma (dashed gray
line) significance. For the 95$\%$ C.L., regions preferred by the
day/night variation data are highlighted in gray.
In the case of 68$\%$ C.L., those regions closely match
the two lower $\Delta m^2_{21}$ 68$\%$ contours shown in the figure.
The black asterisk marks the parameters selected by all solar
neutrino \cite{sno,othersolar,sk4,homestake} and KamLAND data \cite{kamland},
including this work. The SK flux-independent contours agree with those
parameters within two sigma. The previously suggested low (small mixing angle)
solution of neutrino oscillation parameters is excluded at more than four
(five) sigma~\footnote{The Low and small mixing angle solution parameters can 
be seen in~\cite{sno} and~\cite{concha}, respectively.}.

In conclusion, we find an indication of electron-flavor regeneration
in solar neutrino oscillations due to the presence of terrestrial
matter effects. The fit amplitude of the solar zenith angle variation
of the SK solar neutrino interaction rate corresponds to a
day/night asymmetry of $(-3.2\pm1.1(\text{stat})\pm0.5(\text{syst}))\%$,
which deviates from zero by 2.7 $\sigma$. This analysis probes matter
effects directly, since it compares the flavor content of the solar neutrino 
beam with Earth matter to that without. Therefore, this is the first
direct indication that neutrino oscillation probabilities are modified by
the presence of matter. 

\begin{acknowledgments}
  The authors gratefully acknowledge the cooperation of the Kamioka Mining and Smelting Company. Super-K has been built and operated from funds provided by the Japanese Ministry of Education, Culture, Sports, Science and Technology, the U.S. Department of Energy, and the U.S. National Science Foundation. This work was partially supported by the  Research Foundation of Korea (BK21 and KNRC), the Korean Ministry of Science and Technology, the National Science Foundation of China (Grant NO. 11235006), the European Union FP7 ITN INVISIBLES (Marie Curie Actions, PITN-GA-2011-289442) and the State Committee for Scientific Research in Poland.
\end{acknowledgments}

\end{document}

%% file: authors.tex
\newcommand{\AFFicrr}{\affiliation{Kamioka Observatory, Institute for Cosmic Ray Research, University of Tokyo, Kamioka, Gifu 506-1205, Japan}}
\newcommand{\AFFkashiwa}{\affiliation{Research Center for Cosmic Neutrinos, Institute for Cosmic Ray Research, University of Tokyo, Kashiwa, Chiba 277-8582, Japan}}
\newcommand{\AFFipmu}{\affiliation{Kavli Institute for the Physics and
Mathematics of the Universe (WPI), Todai Institutes for Advanced Study,
University of Tokyo, Kashiwa, Chiba 277-8582, Japan }}
\newcommand{\AFFmad}{\affiliation{Department of Theoretical Physics, University Autonoma Madrid, 28049 Madrid, Spain}}
\newcommand{\AFFubc}{\affiliation{Department of Physics and Astronomy, University of British Columbia, Vancouver, BC, V6T1Z4, Canada}}
\newcommand{\AFFbu}{\affiliation{Department of Physics, Boston University, Boston, MA 02215, USA}}
\newcommand{\AFFbnl}{\affiliation{Physics Department, Brookhaven National Laboratory, Upton, NY 11973, USA}}
\newcommand{\AFFuci}{\affiliation{Department of Physics and Astronomy, University of California, Irvine, Irvine, CA 92697-4575, USA }}
\newcommand{\AFFcsu}{\affiliation{Department of Physics, California State University, Dominguez Hills, Carson, CA 90747, USA}}
\newcommand{\AFFcnm}{\affiliation{Department of Physics, Chonnam National University, Kwangju 500-757, Korea}}
\newcommand{\AFFduke}{\affiliation{Department of Physics, Duke University, Durham NC 27708, USA}}
\newcommand{\AFFfukuoka}{\affiliation{Junior College, Fukuoka Institute of Technology, Fukuoka, Fukuoka 811-0295, Japan}}
\newcommand{\AFFgmu}{\affiliation{Department of Physics, George Mason University, Fairfax, VA 22030, USA }}
\newcommand{\AFFgifu}{\affiliation{Department of Physics, Gifu University, Gifu, Gifu 501-1193, Japan}}
\newcommand{\AFFgist}{\affiliation{GIST College, Gwangju Institute of Science and Technology, Gwangju 500-712, Korea}}
\newcommand{\AFFuh}{\affiliation{Department of Physics and Astronomy, University of Hawaii, Honolulu, HI 96822, USA}}
\newcommand{\AFFkanagawa}{\affiliation{Physics Division, Department of Engineering, Kanagawa University, Kanagawa, Yokohama 221-8686, Japan}}
\newcommand{\AFFkek}{\affiliation{High Energy Accelerator Research Organization (KEK), Tsukuba, Ibaraki 305-0801, Japan }}
\newcommand{\AFFkobe}{\affiliation{Department of Physics, Kobe University, Kobe, Hyogo 657-8501, Japan}}
\newcommand{\AFFkyoto}{\affiliation{Department of Physics, Kyoto University, Kyoto, Kyoto 606-8502, Japan}}
\newcommand{\AFFumd}{\affiliation{Department of Physics, University of Maryland, College Park, MD 20742, USA }}
\newcommand{\AFFmit}{\affiliation{Department of Physics, Massachusetts Institute of Technology, Cambridge, MA 02139, USA}}
\newcommand{\AFFmiyagi}{\affiliation{Department of Physics, Miyagi University of Education, Sendai, Miyagi 980-0845, Japan}}
\newcommand{\AFFnagoya}{\affiliation{Solar Terrestrial Environment Laboratory, Nagoya University, Nagoya, Aichi 464-8602, Japan}}
\newcommand{\AFFpol}{\affiliation{National Centre For Nuclear Research, 00-681 Warsaw, Poland}}
\newcommand{\AFFsuny}{\affiliation{Department of Physics and Astronomy, State University of New York at Stony Brook, NY 11794-3800, USA}}
\newcommand{\AFFniigata}{\affiliation{Department of Physics, Niigata University, Niigata, Niigata 950-2181, Japan }}
\newcommand{\AFFokayama}{\affiliation{Department of Physics, Okayama University, Okayama, Okayama 700-8530, Japan }}
\newcommand{\AFFosaka}{\affiliation{Department of Physics, Osaka University, Toyonaka, Osaka 560-0043, Japan}}
\newcommand{\AFFregina}{\affiliation{Department of Physics, University of Regina, 3737 Wascana Parkway, Regina, SK, S4SOA2, Canada}}
\newcommand{\AFFseoul}{\affiliation{Department of Physics, Seoul National University, Seoul 151-742, Korea}}
\newcommand{\AFFshizuokasc}{\affiliation{Department of Informatics in
Social Welfare, Shizuoka University of Welfare, Yaizu, Shizuoka, 425-8611, Japan}}
\newcommand{\AFFskk}{\affiliation{Department of Physics, Sungkyunkwan University, Suwon 440-746, Korea}}
\newcommand{\AFFtohoku}{\affiliation{Research Center for Neutrino Science, Tohoku University, Sendai, Miyagi 980-8578, Japan}}
\newcommand{\AFFtokyo}{\affiliation{The University of Tokyo, Bunkyo, Tokyo 113-0033, Japan }}
\newcommand{\AFFtoront}{\affiliation{Department of Physics, University of Toronto, 60 St., Toronto, Ontario, M5S1A7, Canada }}
\newcommand{\AFFtriumf}{\affiliation{TRIUMF, 4004 Wesbrook Mall, Vancouver, BC, V6T2A3, Canada }}
\newcommand{\AFFtokai}{\affiliation{Department of Physics, Tokai University, Hiratsuka, Kanagawa 259-1292, Japan}}
\newcommand{\AFFtit}{\affiliation{Department of Physics, Tokyo Institute
for Technology, Meguro, Tokyo 152-8551, Japan }}
\newcommand{\AFFtsinghua}{\affiliation{Department of Engineering Physics, Tsinghua University, Beijing, 100084, China}}
\newcommand{\AFFwarsaw}{\affiliation{Institute of Experimental Physics, Warsaw University, 00-681 Warsaw, Poland }}
\newcommand{\AFFuw}{\affiliation{Department of Physics, University of Washington, Seattle, WA 98195-1560, USA}}
\newcommand{\AFFippc}{\affiliation{Institute of Particle Physics, Canada, University of Toronto, 60 Saint George St., Toronta, ON, M5S1A7, Canada}}

\AFFicrr
\AFFkashiwa
\AFFmad
\AFFubc
\AFFbu
\AFFbnl
\AFFuci
\AFFcsu
\AFFcnm
\AFFduke
\AFFfukuoka
\AFFgifu
\AFFgist
\AFFuh
\AFFkek
\AFFkobe
\AFFkyoto
\AFFmiyagi
\AFFnagoya
\AFFsuny
\AFFokayama
\AFFosaka
\AFFregina
\AFFseoul
\AFFshizuokasc
\AFFskk
\AFFtokai
\AFFtokyo
\AFFipmu
\AFFtoront
\AFFippc
\AFFtriumf
\AFFtsinghua
\AFFuw

\author{A.~Renshaw}
\AFFuci
\author{K.~Abe}
\author{Y.~Hayato}
\AFFicrr
\AFFipmu
\author{K.~Iyogi}
\AFFicrr 
\author{J.~Kameda}
\author{Y.~Kishimoto}
\author{M.~Miura} 
\author{S.~Moriyama} 
\author{M.~Nakahata}
\AFFicrr
\AFFipmu 
\author{Y.~Nakano} 
\AFFicrr
% ...

\collaboration{The Super-Kamiokande Collaboration}
\noaffiliation